\documentclass[aps,prb,twocolumn,superscriptaddress,eqsecnum,showpacs,showkeys,amsmath,amssymb]{revtex4}

\usepackage{amsfonts}
\usepackage{amssymb,amsmath}
\usepackage{graphicx}
\usepackage{dcolumn}
\usepackage{bm}
\usepackage{color}

\begin{document}

\title{Magnetocaloric effect in spin-1/2 $XX$ chains with three-spin interactions}

\author{Myroslava Topilko}
\affiliation{Institute for Condensed Matter Physics,
             National Academy of Sciences of Ukraine,
             1 Svientsitskii Street, L'viv-11, 79011, Ukraine}

\author{Taras Krokhmalskii}
\affiliation{Institute for Condensed Matter Physics,
             National Academy of Sciences of Ukraine,
             1 Svientsitskii Street, L'viv-11, 79011, Ukraine}
\affiliation{Department for Theoretical Physics,
             Ivan Franko National University of L'viv,
             12 Drahomanov Street, L'viv-5, 79005, Ukraine}

\author{Oleg Derzhko}
\affiliation{Institute for Condensed Matter Physics,
             National Academy of Sciences of Ukraine,
             1 Svientsitskii Street, L'viv-11, 79011, Ukraine}
\affiliation{Department for Theoretical Physics,
             Ivan Franko National University of L'viv,
             12 Drahomanov Street, L'viv-5, 79005, Ukraine}

\author{Vadim Ohanyan}
\affiliation{Department of Theoretical Physics,
             Yerevan State University,
             Al. Manoogian 1, 0025 Yerevan, Armenia}

\date{\today}

\begin{abstract}
We consider the exactly solvable spin-1/2 $XX$ chain 
with the three-spin interactions of the $XZX+YZY$ and $XZY-YZX$ types
in an external (transverse) magnetic field.
We calculate the entropy and examine the magnetocaloric effect for the quantum spin system.
We discuss a relation between the cooling/heating efficiency and the ground-state phase diagram of the quantum spin model.
We also compare ability to cool/heat in the vicinity of the quantum critical and triple points.
Moreover, 
we examine the magnetocaloric effect 
for the spin-1/2 $XX$ chain with three-spin interactions in a random (Lorentzian) transverse magnetic field.
\end{abstract}



\maketitle

\section{Introduction}
\label{sec1}

In general, 
the magnetocaloric effect (MCE) refers to any change of the temperature of the magnetic material 
under the variation of external magnetic field. 
The revival of interest toward the various aspects of the physics of MCE which has been observed recently 
is mainly connected with the potential room-temperature cooling applications 
(see Refs.~\onlinecite{gschneider, Tishin} for recent reviews). 
Another important application of the MCE is the possibility to map out the $H$-$T$ phase diagram 
by detecting the magnetocaloric anomalies at a magnetic phase transition at high (pulsed) fields. 
For some materials there is no alternative way to do that. 
Since the first successful experiment of adiabatic demagnetization\cite{MacDougall},
MCE is the standard technique of achieving the extremely low temperatures\cite{Strehlow}.

Another important issue of the MCE is its intimate relation with the quantum critical points (QCPs)\cite{sachdev}.
The MCE can be quantified by the adiabatic cooling rate
\begin{eqnarray}
T\Gamma_H
&=&\left( \frac{\partial T}{\partial H}\right)_S
\nonumber\\
&=&-\frac{T}{C_H}\left( \frac{\partial S}{\partial H}\right)_T
=-\frac{T}{C_H}\left( \frac{\partial M}{\partial T}\right)_H,
\label{1.01}
\end{eqnarray}
where $C_H$ is the heat capacity at the constant magnetic field, and $M$ is the magnetization. 
The dependence of the cooling rate on the magnetic field 
is an important characteristic of a specific magnetic material. 
The cooling rate $T\Gamma_H$ is related to the so-called generalized Gr\"{u}neisen ratio,
\begin{eqnarray}
\Gamma_r
=-\frac{1}{T} 
\frac{\left({\partial S}/{\partial r}\right)_T}{\left({\partial S}/{\partial T}\right)_r}
=\frac{1}{T}\left(\frac{\partial T}{\partial r}\right)_S, 
\label{1.02}
\end{eqnarray}
the important quantity characterizing the QCP. 
It is known that the generalized Gr\"{u}neisen ratio changes its sign 
when the parameter $r$ governing the zero-temperature quantum phase transitions crosses its critical value $r_c$, 
i.e., in the QCP\cite{QPT1,QPT2}. 
In the case of MCE $r$ in Eq.~(\ref{1.02}) is the external magnetic field $H$ and QCP corresponds to the critical value $H_c$ 
at which the system undergoes the transition 
between different magnetic structures at zero temperature\cite{Zh_Hon,trippe,PNAS}. 
As the sign of the cooling rate depends on the way magnetic field affects the entropy at isothermal conditions, 
the system can undergo adiabatic cooling as well as adiabatic heating 
under the increasing (or under the decreasing)
of the external magnetic field magnitude. 
Thus, the magnetic materials with complicated structure of zero-temperature (ground-state) phase diagram
display non-trivial MCE with a sequence of cooling and heating.

Very recently, 
exact as well as numerical descriptions of the MCE in various one-dimensional interacting spin systems 
have been attracted much attention\cite{Zh_Hon,trippe,PNAS,JLTP,strecka,derzhko2006,derzhko2007, pereira,strecka2009,hon_wes,ribeiro,ri_mce,Jafari}.
Some two-dimensional systems have been also investigated, 
mainly, in the context of effect of frustration on the MCE\cite{hon2d,shannon2d}. 
The main features of MCE which have been revealed during the investigation of various models are: 
(i) 
essential enhancement of MCE in the vicinity of QCP, 
(ii) 
enhancement of MCE by frustration, 
(iii) 
appearance of the sequence of cooling and heating stages 
during adiabatic (de)magnetization for the systems demonstrating several magnetically ordered ground states,
and 
(iv) 
potential application of MCE data for the investigation of critical properties of the system at hand.

In this paper we continue the investigation of the MCE in one-dimensional quantum spin systems 
admitting the exact solution in the form of free spinless fermions via the Jordan-Wigner transformation 
(see, for instance, Ref.~\onlinecite{tak}). 
Though, the cases of spin-1/2 $XX$ (isotropic) and $XY$ (anisotropic) models 
have been considered in Ref.~\onlinecite{Zh_Hon}, 
there is a series of spin chains with multiple spin interactions introduced by Suzuki in 70s\cite{suzuki1,suzuki2} 
which can be solved by the standard Jordan-Wigner transformation. 
We consider the simplest model of the Suzuki series, 
the spin-1/2 $XX$ chain with three-spin interactions 
both of $XZX+YZY$ and $XZY-YZX$ type\cite{DRD,Gottlieb,Tit_Jap,Lou,Z,krokhmalskii,dksv,G}. 
As has been shown in previous investigations, 
inclusion of the three-spin interactions leads to appearance of new phases in the ground state 
and, thus, to more reach behavior in the vicinity of quantum phase transitions. 
We study two different types of three-spin interactions, 
namely, 
the three-spin interaction of the $XZX+YZY$ type\cite{DRD,Tit_Jap} and of the $XZY-YZX$ type\cite{Gottlieb,Lou}. 
Although these types of three-spin interactions are connected to each other 
by a unitary transformation\cite{krokhmalskii}, 
in the pure form they represent the systems with different symmetries and have different ground-state phase diagrams.
In particular, 
in the former case the ground-state phase diagram contains a point, where three different ground states merge 
[quantum triple point (QTP)].
Appearance of additional parameters in the system, 
the three-spin coupling constants in our case, 
makes possible manipulation of the physical features of MCE, 
namely, 
the position of the QCP and the values of the maximal and minimal temperatures during the adiabatic (de)magnetization. 
The knowledge about manipulation of the MCE physical parameters 
can be very useful for the future quest for the novel magnetic materials and their applications in various aspects. 
On the other hand, 
the appearance of the points
where several magnetically ordered ground states merge on the ground-state phase diagram, 
caused by the inclusion of additional three-spin interactions into the Hamiltonian, 
can lead to essential enhancement of MCE due to large entropy accumulation in such points.
Finally, in real-life materials randomness is always present. 
It can be modeled assuming that on-site fields or intersite interactions acquire random values.
The considered quantum spin chains admit an exact analytical solution for thermodynamics 
in the case when the transverse magnetic field is a random variable with the Lorentzian probability distribution\cite{ddr}.
As a result,
with such a model it is possible to discuss the MCE in the presence of randomness.

The paper is organized as follows. 
At first, we present a general consideration based on the Jordan-Wigner fermionization
(Sec.~\ref{sec2}). 
Next, 
we consider separately  the case of the of $XZX+YZY$ interaction and the case of the $XZY-YZX$ interaction
(Secs.~\ref{sec3} and \ref{sec4}).
After that, 
we consider a random-field spin-1/2 $XX$ chain with three-spin interactions
(Sec.~\ref{sec5}). 
We discuss the MCE in all these cases. 
Finally, we draw some conclusions 
(Sec.~\ref{sec6}).

\section{Jordan-Wigner fermionization and thermodynamic quantities}
\label{sec2}

Let us define the model under consideration.
We consider $N\to\infty$ spins 1/2 placed on a simple chain.
The Hamiltonian of the model looks as follows:
\begin{eqnarray}
\label{2.01} 
{\mathcal{H}} &=& \sum_{n=1}^N
\left[-h s^z_n + J(s^x_ns^x_{n+1} + s^y_ns^y_{n+1}) 
\right.
\nonumber\\
&+& 
K(s^x_ns^z_{n+1}s^x_{n+2} + s^y_ns^z_{n+1}s^y_{n+2})
\nonumber\\
&+& 
\left.
E(s^x_ns^z_{n+1}s^y_{n+2} - s^y_ns^z_{n+1}s^x_{n+2})
\right],
\end{eqnarray}
where $h$ is the external (transverse) magnetic field,
$J$ is the isotropic $XY$ (i.e., $XX$) exchange interaction constant
(in what follows we will set $J=1$ to fix the units),
and
$K$ and $E$ are the constants of the two types of three-spin exchange interactions.
We imply periodic boundary conditions in Eq.~(\ref{2.01}) for convenience.

The Hamiltonian (\ref{2.01}) can be brought to the diagonal Fermi-form 
after applying at first the Jordan-Wigner transformation to spinless fermions,
\begin{eqnarray}
\label{2.02}
s^+_n &=& s^x_n + is^y_n = P_{n-1}c^\dag_n,
\;
s^-_n = s^x_n - is^y_n = P_{n-1}c_n,
\nonumber\\
c^\dag_n &=& P_{n-1}s^+_n,
\;
c_n = P_{n-1}s^-_n,
\nonumber\\
P_m &=& \prod^m_{j=1}(1-2c^\dag_jc_j) = \prod^m_{j=1}(-2s_j^z),
\end{eqnarray}
and performing further the Fourier transformation,
\begin{eqnarray}
\label{2.03}
c^\dag_n &=& \frac{1}{\sqrt{N}} \sum_k e^{ikn}c_k^\dag,
\;
c_n = \frac{1}{\sqrt{N}} \sum_k e^{-ikn}c_k,
\nonumber\\
c^\dag_k &=& \frac{1}{\sqrt{N}} \sum_{n=1}^N e^{-ikn}c_n^\dag,
\;
c_k = \frac{1}{\sqrt{N}} \sum_{n=1}^N e^{ikn}c_n,
\end{eqnarray}
$k= 2\pi m/N$, $m = -N/2,\ldots,N/2-1$
(we assume that $N$ is even without loss of generality).
As a result,
\begin{eqnarray}
\label{2.04}
{\cal{H}} &=& \sum_k\varepsilon_k\left(c_k^\dag c_k-\frac{1}{2}\right),
\nonumber\\
\varepsilon_k &=& -h+J\cos k-\frac{K}{2}\cos(2k)-\frac{E}{2}\sin(2k).
\end{eqnarray}

Using Eq.~(\ref{2.04}) we can easily calculate the partition function for the spin model (\ref{2.01})
\begin{eqnarray}
\label{2.05}
Z(T,h,N)={\rm{Tr}}e^{-{\cal{H}}/T}=\prod_k2{\rm{ch}}\frac{\varepsilon_k}{2T}
\end{eqnarray}
(we set $k_{\rm{B}}=1$).
Various thermodynamic quantities,
such as the Helmholtz free energy, the entropy, and the specific heat (per site)
immediately follow from Eq.~(\ref{2.05}):
\begin{eqnarray}
\label{2.06}
f(T,h) &=& -\lim_{N\rightarrow\infty}\frac{T\ln Z(T,h,N)}N
\nonumber\\
&=&\frac1{2\pi}\int_{-\pi}^\pi dk \left(\frac{\varepsilon_k}2+T\ln n_k\right),
\nonumber\\
s(T,h)&=&-\frac{\partial f(T,h)}{\partial T}
\nonumber\\
&=& -\frac1{2\pi}\int_{-\pi}^\pi dk\left(\ln{n_k}+\frac{\varepsilon_k}Te^{\varepsilon_k/T}n_k\right),
\nonumber\\
c(T,h)&=&T\frac{\partial s(T,h)}{\partial T}
\nonumber\\
&=&\frac1{2\pi T^2}\int_{-\pi}^\pi dk\varepsilon_k^2n_k(1-n_k);
\end{eqnarray}
here $n_k=1/(e^{\varepsilon_k/T}+1)$ are the occupation numbers of spinless fermions.
Furthermore,
we get
\begin{eqnarray}
\label{2.07}
m(T,h)&=&\lim_{N\to\infty}\frac{1}{N}\sum_{n=1}^N\langle s_n^z\rangle
=-\frac{\partial f(T,h)}{\partial h} 
\nonumber\\
&=&\frac{1}{2\pi}\int_{-\pi}^{\pi}dk \left(n_k-\frac{1}{2}\right), 
\nonumber\\
\frac{\partial m(T,h)}{\partial T}
&=&\frac{1}{2\pi T^2}\int_{-\pi}^{\pi}dk \varepsilon_k n_k (1-n_k),
\nonumber\\
\Gamma_h
&=&-\frac{1}{c(T,h)}\frac{\partial m(T,h)}{\partial T}
\nonumber\\
&=&-\frac{\int_{-\pi}^{\pi}dk \varepsilon_k n_k(1-n_k)}{\int_{-\pi}^{\pi}dk \varepsilon_k^2 n_k (1-n_k)}.
\end{eqnarray}

It may be useful to rewrite the formulas for thermodynamic quantities (\ref{2.06}), (\ref{2.07})
in terms of the density of states
\begin{eqnarray}
\rho(\omega)=\lim_{N\to\infty}\frac{1}{N}\sum_k\delta(\omega-\varepsilon_k)
=\frac{1}{2\pi}\int_{-\pi}^{\pi}dk\delta(\omega-\varepsilon_k).
\label{2.08}
\end{eqnarray}
We have
\begin{eqnarray}
\label{2.09} 
f(T,h)&=-T&\int_{-\infty}^{\infty} d\omega \rho(\omega)
\ln\left( 2{\rm{ch}}\frac{\omega}{2T}\right),
\nonumber\\
s(T,h)&=&\int_{-\infty}^{\infty} d\omega \rho(\omega)
\left[ \ln\left(2{\rm{ch}}\frac{\omega}{2T}\right) 
-
\frac{\omega}{2T}{\rm{th}}\frac{\omega}{2T} \right],
\nonumber\\
c(T,h)&=&\frac{1}{4T^2}\int_{-\infty}^{\infty} d\omega \rho(\omega)
\frac{\omega^2}{{\rm{ch}}^2[{\omega}/(2T)]},
\nonumber\\
m(T,h)&=&-\frac{1}{2}\int_{-\infty}^{\infty} d\omega \rho(\omega){\rm{th}}\frac{\omega}{2T},
\nonumber\\
\frac{\partial m(T,h)}{\partial T}
&=&\frac{1}{4T^2}\int_{-\infty}^{\infty} d\omega \rho(\omega)\frac{\omega}{{\rm{ch}}^2[{\omega}/(2T)]},
\nonumber\\
\Gamma_h
&=&
-\frac{\int_{-\infty}^{\infty} d\omega \rho(\omega){\omega}/{\{{\rm{ch}}^2[{\omega}/(2T)]}\}}
{\int_{-\infty}^{\infty} d\omega \rho(\omega){\omega^2}/{\{{\rm{ch}}^2[{\omega}/(2T)]}\}}.
\end{eqnarray}
Formulas (\ref{2.09}) are extremely useful for consideration of the random  spin-1/2 $XX$ chains 
within the Green's functions approach\cite{gf,ddr},
since that method permits to calculate the random-averaged density of states (\ref{2.08}),
see Ref.~\onlinecite{ddr} and Sec.~\ref{sec5}.

Although the presented above formulas give a comprehensive description of the quantum spin system (\ref{2.01})
(and, in particular, the MCE), 
the thermodynamic behavior is somehow hidden behind one-fold integrals in Eqs.~(\ref{2.06}), (\ref{2.07}).
More explicit dependencies of thermodynamic quantities on temperature and field
can be derived, e.g., in the low-temperature limit.
Let us briefly discuss what happens with Eqs.~(\ref{2.06}), (\ref{2.07}) when $T\to 0$,
see also Sec.~\ref{sec3}.
We note that $n_k(1-n_k)=1/\{4{\rm{ch}}^2[\varepsilon_k/(2T)]\}$
and therefore as $T\to 0$
only a small region 
where $\varepsilon_k\approx 0$ is relevant in the integrals yielding $c(T,h)$ or $\partial m(T,h)/\partial T$
in the low-temperature limit.
Clearly, 
if the energy spectrum of spinless fermions is gapped we immediately get 
that $c(T,h)$ and $\partial m(T,h)/\partial T$ vanishes as $T\to 0$.
We turn to the case of a gapless energy spectrum of spinless fermions.
Assume that we have
$\varepsilon_k=\varepsilon_i^{(z)}(k-k_i)^z/z!+\ldots$
around $k_i$ satisfying $\varepsilon_{k_{i}}=0$.
Then we immediately find that $c(T,h)\propto T^{1/z}$.
Note also that $s(T,h)\propto T^{1/z}$ 
and in a ``flat-band-like'' limit $z\to\infty$ 
the entropy becomes independent on temperature
(for a discussion of true flat-band spin systems see Refs.~\onlinecite{flatband,Zh_Hon,derzhko2006,derzhko2007}).
While estimating $\partial m(T,h)/\partial T$ for odd $z$ (e.g., for $z=1$) 
we have to take higher-order terms in the expansion of $\varepsilon_k$ around $k=k_i$.
For even $z$ 
[$z=2$ for the QCP and $z=4$ for the QTP, see Eq.~(\ref{3.01})]
we get 
$\partial m(T,h)/\partial T \propto T^{1/z-1}$
and therefore
$\Gamma_h\propto T^{-1}$.

Alternatively the critical behavior can be derived using formulas (\ref{2.09}).
The factor $1/{\rm{ch}}^2[\omega/(2T)]$ in the integrands for $c(T,h)$ and $\partial m(T,h)/\partial T$ implies 
that in the limit $T\to 0$ only a small region where $\omega\approx 0$ is relevant.
Around the QCP $\rho(\omega)\propto\omega^{-1/2}$
and therefore 
$c(T,h)\propto T^{1/2}$,
$\partial m(T,h)/\partial T\propto T$,
whereas
around the QTP $\rho(\omega)\propto\omega^{-3/4}$ 
(see Ref.~\onlinecite{krokhmalskii})
and as a result 
$c(T,h)\propto T^{1/4}$,
$\partial m(T,h)/\partial T\propto T^{-3/4}$.

In the case of randomness considered in Sec.~\ref{sec5},
van Hove peculiarities in the density of states are smeared out,
$\overline{\rho(\omega)}$ has a finite nonzero value for any $\omega$
and, in particular, $\overline{\rho(\omega)}=\overline{\rho(0)}+\ldots$ around $\omega=0$.
As a result,
$\overline{c(T,h)}\propto T$
for a sufficiently low temperature $T\to 0$
[an estimate of $\partial \overline{m(T,h)}/\partial T$ 
requires higher-order terms in the expansion of $\overline{\rho(\omega)}$ around $\omega=0$].
The boundaries between different ground-state phases disappear
since the quantum phase transition transforms into a crossover\cite{ddr}.

\section{Three-spin interactions of $XZX+YZY$ type}
\label{sec3}

\begin{figure}
\begin{center}
\includegraphics [width=8cm]{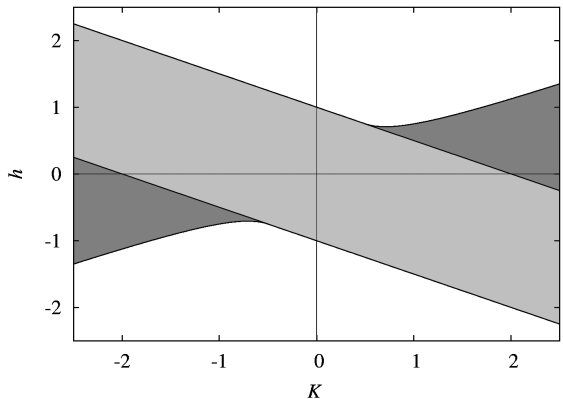}
\caption{
The ground-state phase diagram in the $K-h$ plane of the model (\ref{2.01}) with $J=1$ and $E=0$. 
The dark-gray regions correspond to the spin-liquid II phase,
the light-gray region corresponds to the spin-liquid I phase, 
and the white regions correspond to the ferromagnetic phase.
The lines $h^{\star}(K)$ which separate different regions 
correspond to quantum phase transitions between different ground-state phases.}
\label{fig1}
\end{center}
\end{figure}

Now we consider the chain with two-spin interactions and three-spin interactions of $XZX + YZY$ type\cite{Tit_Jap, DRD}.
In this case we put in Eq.~(\ref{2.01}) $E=0$
and the energy spectrum of spinless fermions in Eq.~(\ref{2.04}) reads:
\begin{eqnarray}
\label{3.01}
\varepsilon_k=-h+J\cos k-\frac{K}{2}\cos(2k).
\end{eqnarray}
The third term in (\ref{3.01}) may lead to a new ground-state phase,
the so-called spin-liquid II phase\cite{Tit_Jap}.
The phase diagram in the ground state is shown in Fig.~\ref{fig1}.
While $|K|$ is less than 1/2 
($J=1$)
there are only two ground-state phases:
the spin-liquid I phase and the ferromagnetic phase.
However, when $|K|>1/2$ there is in addition one more ground-state phase,
the spin-liquid II phase.
It is worthy noting that there are two special points
($K=1/2$, $h=3/4$ and $K=-1/2$, $h=-3/4$)
on the ground-state phase diagram
at which all three ground-state phases meet (QTPs).
For further details see Refs.~\onlinecite{Tit_Jap,krokhmalskii}.

The entropy of the spin system is a function of the temperature $T$, the magnetic field $h$, and of the parameter $K$
[see Eq.~(\ref{2.06}) in which we have to use $\varepsilon_k$ given in Eq.~(\ref{3.01}) ($J=1$)].

\begin{figure}
\includegraphics [width=8cm]{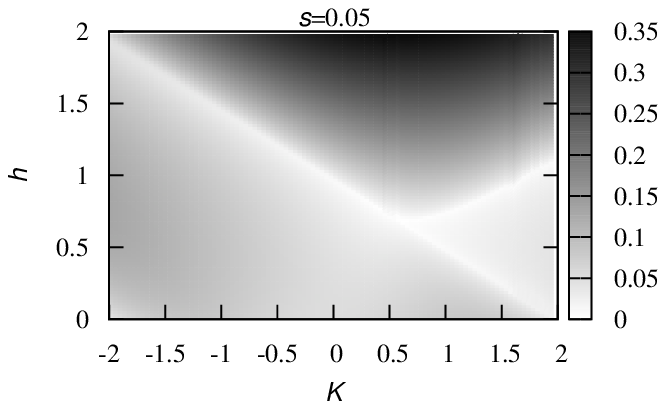}\\
\vspace{3mm}
\includegraphics [width=7cm]{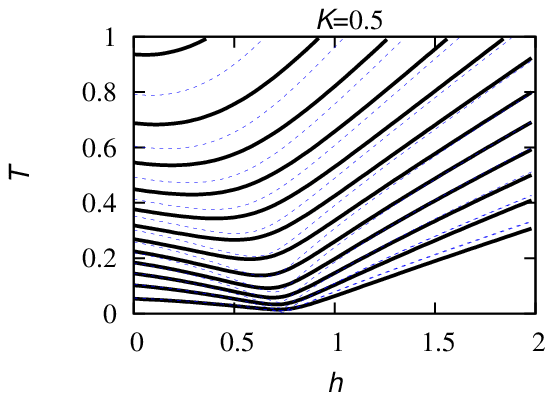}\\
\vspace{3mm}
\includegraphics [width=7cm]{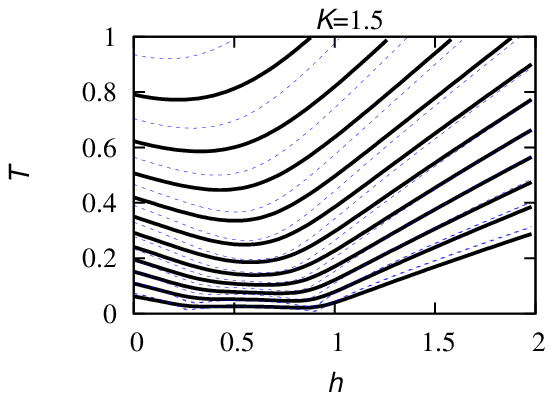}
\caption{
Upper panel:
Grayscale plots of the temperature as it follows from the condition $s(T,h)=0.05$ 
in the $K-h$ plane for model (\ref{2.01}) with $J=1$ and $E=0$. 
Two lower panels:
Isentropic dependence $T$ vs $h$ at $s=0.05,0.10,\ldots,0.60$ (from bottom to top in each panel) 
for model (\ref{2.01}) with $J=1$, $K=0.5,\;1.5$, $E=0$.
Thin broken lines correspond to the nonrandom model,
thick solid lines correspond to the random-field model (\ref{5.01}) with $\Gamma=0.1$.}
\label{fig2}
\end{figure}

Now we turn to a discussion of the MCE in its classical interpretation
as an adiabatic change of the temperature of the considered model under field variation.
Grayscale plots of the temperature at constant entropy $s(T,h)=0.05$ in the $K-h$ plane
are shown in the upper panel of Fig.~\ref{fig2}.
The lowest values of the temperature are around the QTP $K=1/2$, $h=3/4$.
Two lower panels of Fig.~\ref{fig2} supplement the upper one.
In two lower panels of Fig.~\ref{fig2} we show by thin broken lines
the dependencies $T(h)$ at fixed values of $s = 0.05, 0.10,\ldots,0.60$ 
for spin model (\ref{2.01}) with $J=1$, $K=0.5,\;1.5$, $E=0$.

\begin{figure}
\includegraphics [width=8cm]{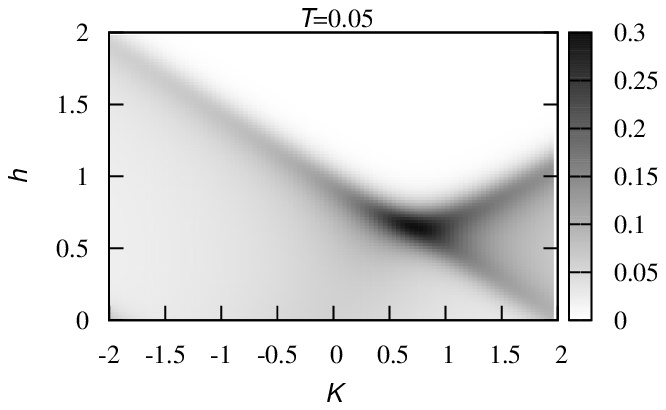}\\
\vspace{3mm}
\includegraphics [width=7cm]{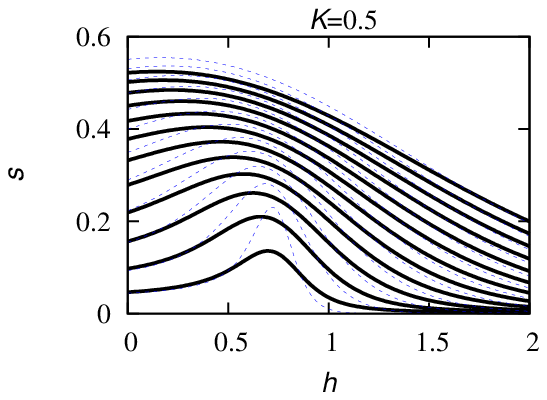}\\
\vspace{3mm}
\includegraphics [width=7cm]{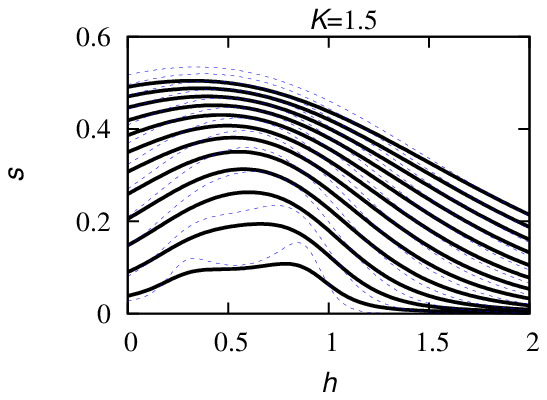}
\caption{
Upper panel:
Grayscale plots of the entropy $s(T,h)$ in the $K-h$ plane at $T=0.05$ for model (\ref{2.01}) with $J=1$ and $E=0$.
Two lower panels:
Isothermal dependence $s$ vs $h$ at $T=0.05,0.10,\ldots,0.60$ (from bottom to top in each panel) 
for model (\ref{2.01}) with $J=1$, $K=0.5,\;1.5$, $E=0$.
Thin broken lines correspond to the nonrandom model,
thick solid lines correspond to the random-field model (\ref{5.01}) with $\Gamma=0.1$.}
\label{fig3}
\end{figure}

According to Eq.~(\ref{1.01}),
to discuss the MCE
we may analyze alternatively an isothermal change of the entropy under field variation.
In the upper panel of Fig.~\ref{fig3} we show grayscale plots of the entropy in the $K-h$ plane for constant temperature $T=0.05$.
Again two lower panels of Fig.~\ref{fig3} supplement the upper one. 
In two lower panels of Fig.~\ref{fig3} we show by thin broken lines
the dependencies $s(h)$ at $T=0.05,\;0.10,\ldots,0.60$
for spin model (\ref{2.01}) with $J=1$, $K=0.5,\;1.5$, $E=0$.

Comparing Fig.~\ref{fig2} and Fig.~\ref{fig3} with the ground-state phase diagram in Fig.~\ref{fig1}
one can note that the MCE in the low-$s$ or low-$T$ regimes perfectly reproduces
the ground-state phase transition lines.

Furthermore,
consider, e.g., $K=1/2$ 
(thin broken lines in the middle panel in Fig.~\ref{fig2}). 
Clearly if we decrease adiabatically $h$ from 2 to 3/4 
the temperature noticeably falls down
(e.g., approximately from 0.3388 to 0.0002 at $s=0.05$ or from 0.4331 to 0.0025 at $s=0.10$).
We turn to the results shown by thin broken lines in the middle panel in Fig.~\ref{fig3}.
If we decrease isothermally $h$ from 2 to 3/4, 
the entropy of spin system noticeably increases
(e.g., approximately from 0.00000 to 0.2195 at $T=0.05$ or from 0.00001 to 0.2670 at $T=0.10$) 
meaning that the spin system absorbs from the thermostat the heat $T\left[s(h=3/4)-s(h=2)\right]$ 
(that is,
$\approx 0.0110$ 
at $T=0.05$ 
or 
$\approx 0.0267$ 
at $T=0.10$)
per site.

\begin{figure}
\includegraphics [width=7cm]{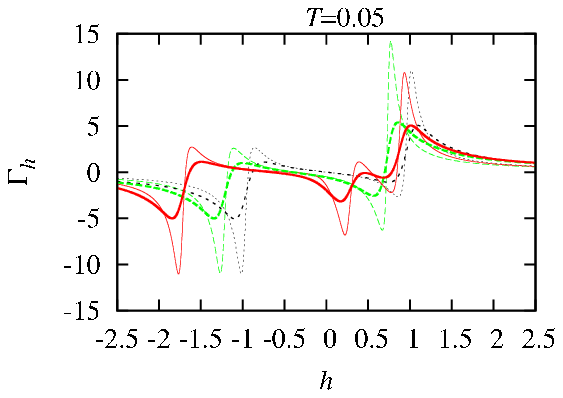}
\caption{
$\Gamma_h$ vs $h$ 
for model (\ref{2.01}) with $J=1$, $K=0$ (dotted), $K=0.5$ (dashed), $K=1.5$ (solid), and $E=0$ at $T=0.05$.
Thin lines correspond to the nonrandom model,
thick lines correspond to the random-field model (\ref{5.01}) with $\Gamma=0.1$.}
\label{fig4}
\end{figure}

In Fig.~\ref{fig4} we plot by thin lines the dependence $\Gamma_h(h)$ at $T=0.05$
for a few values of $K=0$ (dotted), $K=0.5$ (dashed), and $K=1.5$ (solid).
From the first glance it becomes clear 
that one peak (around $h\approx 0.75$) is higher than others.
It might be interesting to compare the height of maxima in the dependence $\Gamma_h(h)$ for $T=0.05$,
in particular the height of the high-field ones which correspond to cooling while $h$ decreases starting from high fields
(see thin lines in Fig.~\ref{fig4} for $h=0.75\ldots 2$).
For $K=0,\;0.5,\;1.5$ we have
$\Gamma_h(h)\approx 11.00,\;14.22,\;10.80$ 
at 
$h\approx 1.02,\;0.77,\;0.93$, respectively.
(We recall here that the quantum phase transition occurs at $h^\star=1,\;3/4,\;11/12$ 
for $K=0,\;0.5,\;1.5$, respectively, 
see Fig.~\ref{fig1} and Ref.~\onlinecite{krokhmalskii}.) 
Clearly, the value of the cooling rate around the QTP is about 130\% of such a value around the QCP 
(e.g., for the studied earlier $K=0$ case\cite{Zh_Hon}).
At lower temperatures the heights of maxima increase 
(e.g., 
$\Gamma_h(h)\approx 55.56,\;73.04,\;55.41$ 
at 
$h\approx 1.003,\;0.754,\;0.920$
for $T=0.01$
and
$\Gamma_h(h)\approx 556.51,\;739.48,\;556.61$ 
at 
$h\approx 1.0003,\;0.7504,\;0.9170$
for $T=0.001$)
but the relation between the heights changes only very slightly and approaches roughly 3:4:3.
Thus, one gets about 33\% larger change in temperature for the same adiabatic change of the field 
being performed around QTP in comparison with being performed around QCP.

This can be illustrated further
while considering Eq.~(\ref{2.07}) in the limit $T\to 0$.
We recall that according to Eq.~(\ref{3.01})
$\varepsilon_k=-h+J-K/2-(J/2-K)k^2
+(J-8K)k^4/4!+\ldots$
and therefore while approaching a high-field peculiar point $h^\star$ from above,
i.e., by decreasing $h$, $h\to h^\star+0$ 
(a ferromagnetic--to--spin-liquid transition),
we have
\begin{eqnarray}
\varepsilon_k=-(h-h^\star)-\frac{1}{2}k^2+\frac{1}{4!}k^4+\ldots
\label{3.02}
\end{eqnarray}
for $K=0$ (QCP) with $h^\star=1$ and 
\begin{eqnarray}
\varepsilon_k=-(h-h^\star)-\frac{1}{8}k^4+\frac{1}{48}k^6 -\frac{1}{640}k^8+\ldots
\label{3.03}
\end{eqnarray}
for $K=1/2$ (QTP) with $h^\star=3/4$.
Moreover, we may single out two regimes.
In the first regime, 
we first take the limit $T\to 0$ and then $h-h^\star\to +0$ [i.e., $(h-h^\star)/T\gg 1$].
In the second regime, 
we first put $h-h^\star=0$ and then take the limit $T\to 0$ [i.e., $(h-h^\star)/T\ll 1$].
[For the high-field peaks in Fig.~\ref{fig4} we have $(h-h^\star)/T\approx 0.3\ldots 0.4$.]

If $T\to 0$ and $(h-h^\star)/T\gg 1$ we can write
$n_k(1-n_k)\approx e^{-\vert\varepsilon_k\vert/T}$
and hence
\begin{eqnarray}
\label{3.04}
n_k(1-n_k)\propto e^{-k^2/(2T)}
\end{eqnarray}
for $K=0$ and
\begin{eqnarray}
\label{3.05}
n_k(1-n_k)\propto e^{-k^4/(8T)}
\end{eqnarray}
for $K=1/2$,
see Eqs.~(\ref{3.02}) and (\ref{3.03}).
Furthermore, 
we can extend the limits of integration with respect to $k$ 
in two relevant integrals in the formula for $\Gamma_h$ (\ref{2.07})
to $-\infty$ and $\infty$.
Using Eqs.~(\ref{3.02}), (\ref{3.03}), (\ref{3.04}), and (\ref{3.05})
we immediately find that $\Gamma_h\to 1/(h-h^{\star})$ as $h\to h^\star+0$ in the limit $T=0$ 
for both cases, $K=0$ and $K=1/2$.
For small but finite $T$ we obtain different results for $K=0$ and $K=1/2$.
Although all relevant integrals are doable 
with the help of the well-known formula for the gamma function $\Gamma(z)$ 
(see, e.g., Ref.~\onlinecite{fedoryuk})
\begin{eqnarray}
\label{3.06}
\int_0^{\infty}dxx^{\beta-1}e^{-\lambda x^{\alpha}}
=\frac{1}{\alpha}\lambda^{-\beta/\alpha}\Gamma\left(\frac{\beta}{\alpha}\right),
\end{eqnarray}
$\Re\lambda>0$, $\alpha,\;\beta>0$,
it is simpler to find required results by using MAPLE codes.
Namely,
for the case $K=0$ we have
\begin{eqnarray}
\label{3.07}
\Gamma_h\approx
\frac{1}{h-h^\star}
\frac{1+\epsilon/2-\epsilon T/8}
{1+\epsilon -(h-4)\epsilon^2/4}
\end{eqnarray}
with $\epsilon=T/(h-h^\star)$, $h^\star=1$,
whereas for the case $K=1/2$ we have
\begin{eqnarray}
\label{3.08}
\Gamma_h\approx
\frac{1}{h-h^\star}
\frac{1+\epsilon /4 - 0.119 \epsilon \sqrt{T} +\epsilon T/32}
{1+\epsilon/2 - 0.239 \epsilon \sqrt{T} + (h+17/4)\epsilon^2/16}
\end{eqnarray}
with $\epsilon=T/(h-h^\star)$, $h^\star=3/4$.
Approximate analytical formulas (\ref{3.07}) and (\ref{3.08}) yield $1:1.39$
for ratio of the heights of peaks in the dependence $\Gamma_h(h)$ around QCP and QTP at $T=0.05$ 
that is in a reasonable agreement with exact numerical calculation according to Eq.~(\ref{2.07}).

If we put at first $h=h^\star$ and then assume $T\to 0$ we can write
\begin{eqnarray}
\label{3.09}
n_k(1-n_k)=\frac{1}{2+e^{\varepsilon_k/T}+e^{-\varepsilon_k/T}}
\nonumber\\
\approx 
\frac{1}{4+(ak^z)^2/T^2}
\approx
\frac{1}{4}e^{-a^2k^{2z}/(4T^2)}
\end{eqnarray}
with $\varepsilon_k=-ak^z<0$, $z=2$ for the QCP and $z=4$ for the QTP,
see Eqs.~(\ref{3.02}) and (\ref{3.03}).
Again the limits of integration with respect to $k$ in the two relevant integrals in Eq.~(\ref{2.07})
can be extended to $-\infty$ and $\infty$.
After simple calculations using (\ref{3.06}) we find 
that $\Gamma_h$ at QTP relates to $\Gamma_h$ at QCP as 
$\Gamma(5/8)/\Gamma(9/8)$ to $\Gamma(3/4)/\Gamma(5/4)$,
i.e., as 1.1267\ldots :1.
Interestingly,
for arbitrary $z$ we have 
$\Gamma_h\to \{\Gamma[(z+1)/(2z)]/\Gamma[(2z+1)/(2z)]\}/(2T)\to\sqrt{\pi}/(2T)$ if $z\to\infty$.
Thus in such a case ($z\to\infty$) $\Gamma_h$ is about 131\% of $\Gamma_h$ at the QCP with $z=2$.

In summary, 
as can be seen from numerical calculations of $\Gamma_h$ (\ref{2.07}) reported in Fig.~\ref{fig4}
as well as from analytical considerations in specific limits\cite{footnote}, 
the efficiency of cooling while decreasing $h$ starting from the high-field limit 
is higher around the QTP than around the QCP.
We note here
that enhancement of the MCE in the frustrated $J_1-J_2$ antiferromagnetic Heisenberg chain 
due to a cancellation of the leading $k^2$-term in the one-particle energy spectrum at the point $J_2=J_1/4$ 
was discussed in Ref.~\onlinecite{Zh_Hon}.
For the case at hand (\ref{2.01}), (\ref{3.01}),
the softening in the excitation spectrum occurs at the QTP 
which emerges owing to the three-spin interactions of $XZX+YZY$ type.

\section{Three-spin interactions of $XZY-YZX$ type}
\label{sec4}

Now we turn to the spin-1/2 $XX$ chain with three-spin interactions of $XZY-YZX$ type\cite{Lou, Gottlieb}.
In this case we put $K=0$ in Eq.~(\ref{2.01}) 
and the energy spectrum of spinless fermions in Eq.~(\ref{2.04}) reads:
\begin{eqnarray}
\label{4.01}
\varepsilon_k =-h+J\cos k-\frac{E}{2}\sin(2k).
\end{eqnarray}

\begin{figure}
\begin{center}
\includegraphics [width=8cm]{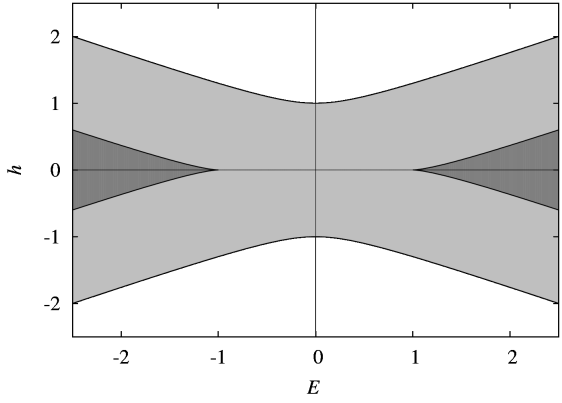} 
\caption{
The ground-state phase diagram in the $E-h$ plane of the model (\ref{2.01}) with $J=1$ and $K=0$.
The dark-gray regions correspond to the spin-liquid II phase,
the light-gray region corresponds to the spin-liquid I phase,
and the white regions correspond to the ferromagnetic phase.
The lines $h^{\star}(E)$ which separate different regions 
correspond to quantum phase transitions between different ground-state phases.}
\label{fig5}
\end{center}
\end{figure}

The ground-state phase diagram of the model is shown in Fig.~\ref{fig5}.
This type of three-spin interactions leads to the spin-liquid II phase too.
But for the model  (\ref{4.01}), in contrast to the model (\ref{3.01}),
we do not have a QTP
and all phase transition lines separate two different phases only,
compare Fig.~\ref{fig5} and Fig.~\ref{fig1}.
For further details see Refs.~\onlinecite{Gottlieb,Lou}.

\begin{figure}
\includegraphics [width=8cm]{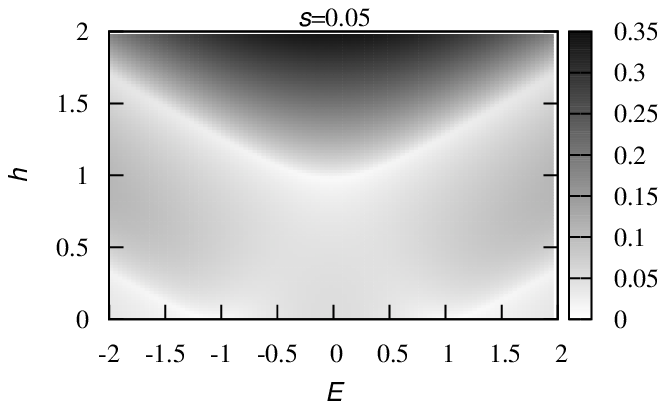}\\
\vspace{3mm}
\includegraphics [width=7cm]{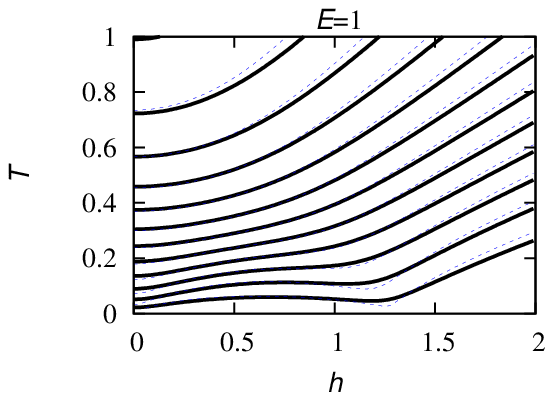}\\
\vspace{3mm}
\includegraphics [width=7cm]{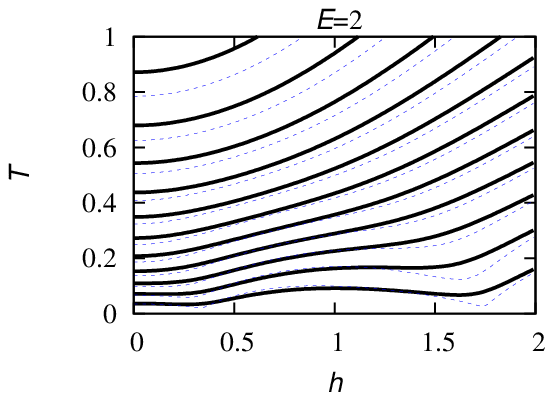}
\caption{
Upper panel:
Grayscale plots of the temperature
as it follows from the condition $s(T,h)=0.05$ in the $E-h$ plane for model (\ref{2.01}) with $J=1$ and $K=0$.
Two lower panels:
Isentropic dependence $T$ vs $h$ at $s=0.05,0.10,\ldots,0.60$ (from bottom to top in each panel) 
for model (\ref{2.01}) with $J=1$, $K=0$, $E=1,\;2$.
Thin broken lines correspond to the nonrandom model,
thick solid lines correspond to the random-field model (\ref{5.01}) with $\Gamma=0.1$.}
\label{fig6}
\end{figure}

\begin{figure}
\includegraphics [width=8cm]{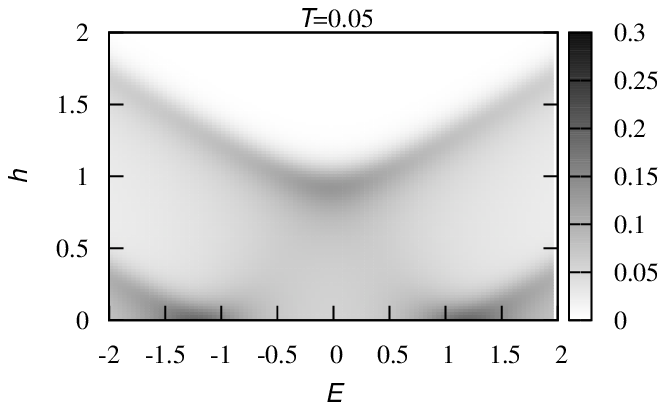}\\
\vspace{3mm}
\includegraphics [width=7cm]{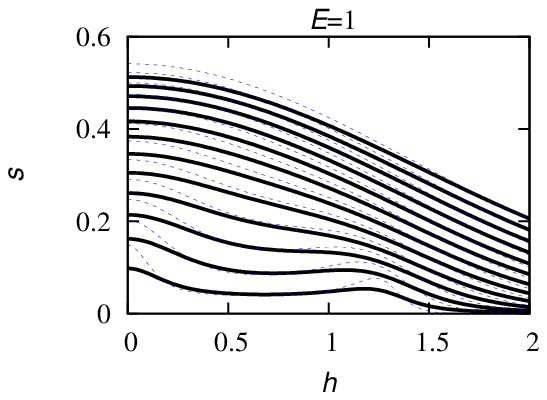}\\
\vspace{3mm}
\includegraphics [width=7cm]{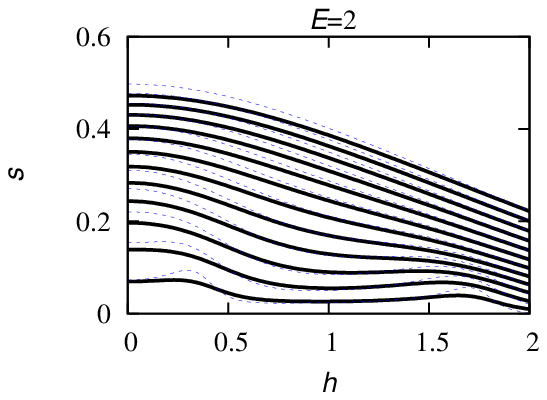}
\caption{
Upper panel:
Grayscale plots of the entropy $s(T,h)$ in the $E-h$ plane at $T=0.05$ for model (\ref{2.01}) with $J=1$ and $K=0$.
Two lower panels:
Isothermal dependencies $s$ vs $h$ at $T=0.05,0.10,\ldots ,0.60$ (from bottom to top in each panel) 
for model (\ref{2.01}) with $J=1$, $K=0$, $E=1,\;2$.
Thin broken lines correspond to the nonrandom model,
thick solid lines correspond to the random-field model (\ref{5.01}) with $\Gamma=0.1$.}
\label{fig7}
\end{figure}

\begin{figure}
\includegraphics [width=7cm]{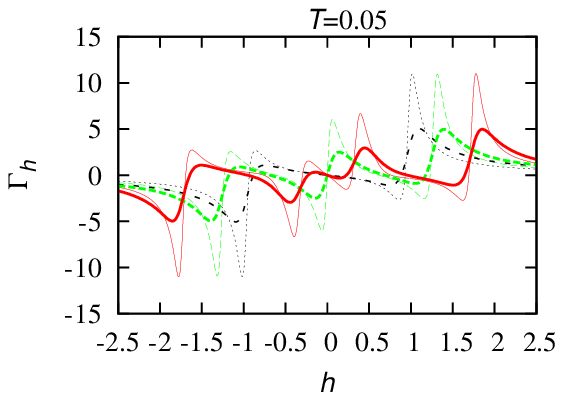}
\caption{
$\Gamma_h$ vs $h$ 
for model (\ref{2.01}) with $J=1$, $K=0$, $E=0$ (dotted), $E=1$ (dashed), and $E=2$ (solid) at $T=0.05$.
Thin lines correspond to the nonrandom model,
thick lines correspond to the random-field model (\ref{5.01}) with $\Gamma=0.1$.}
\label{fig8}
\end{figure}

Figs.~\ref{fig6}, \ref{fig7}, and \ref{fig8}
are similar to
Figs.~\ref{fig2}, \ref{fig3}, and \ref{fig4}.
Again the MCE in the low-$s$ and low-$T$ regimes (Figs.~\ref{fig6}, \ref{fig7}, \ref{fig8})
indicates the ground-state phase transition lines seen in Fig.~\ref{fig5}.
Again we observe that cooling/heating is especially efficient around QCPs.
Since the model with three-spin interactions of $XZY-YZX$ type does not have a QTP,
in all cases $\Gamma_h$ behaves like it should around QCP,
see thin lines in Fig.~\ref{fig8}.
For example, 
the height of the high-field peaks 
in the dependence $\Gamma_h(h)$ at $T=0.05$  for various $E$ is the same
(see thin lines for $h=1\ldots 2$ in Fig.~\ref{fig8}).


\section{Random (Lorentzian) transverse magnetic field}
\label{sec5}

In this section we use recent results 
on thermodynamics of the spin-1/2 $XX$ chain with three-spin interactions in a random transverse field\cite{ddr} 
to discuss the influence of randomness on the MCE.
To be specific, 
we consider the Hamiltonian (\ref{2.01}) and make the change $h\to h_n$,
where $h_n$ is the random transverse magnetic field with the Lorentzian probability distribution
\begin{eqnarray}
\label{5.01}
p(h_n)=\frac{1}{\pi}\frac{\Gamma}{(h_n-h)^2+\Gamma^2}.
\end{eqnarray}
Now $h$ is the mean value of $h_n$ 
and the parameter $\Gamma$ controls the strength of the Lorentzian disorder.
The nonrandom case can be reproduced if $\Gamma$ is sent to 0.
All (random-averaged) thermodynamic quantities of the random quantum spin system (\ref{2.01}), (\ref{5.01})
can be expressed through the (random-averaged) density of states\cite{ddr}. 
For instance, 
in the formula for the entropy in Eq.~(\ref{2.09}) 
we now have to use the density of states 
\begin{eqnarray}
\label{5.02} 
\overline{\rho(\omega)} &=& \mp\frac{1}{\pi}\Im \overline{G_{jj}^{\mp}(\omega)},
\nonumber\\
\overline{G_{jj}^{\mp}(\omega)} 
&=& \frac{4}{K} 
\left[\frac{z_1}{(z_1-z_2)(z_1-z_3)(z_1-z_4)} 
\right.
\nonumber\\
&+&
\left. 
\frac{z_2}{(z_2-z_1)(z_2-z_3)(z_2-z_4)} \right]
\end{eqnarray}
and 
$\vert z_1\vert\le \vert z_2\vert\le \vert z_3\vert\le \vert z_4\vert$ 
are the solutions of a certain quartic equation, see Ref.~\onlinecite{ddr}.
In the case $E=0$ the relevant quartic equation, 
\begin{eqnarray}
\label{5.03}
z^4-\frac{2J}{K}z^3+\frac{4}{K}(\omega+h\pm i\Gamma)z^2-\frac{2J}{K}z+1=0,
\end{eqnarray}
can be reduced to the quadratic one
that further simplifies calculations.
In the case $K=0$ the relevant quartic equation, 
\begin{eqnarray}
\label{5.04}
z^4-\frac{2iJ}{E}z^3+\frac{4i}{E}(\omega+h\pm i\Gamma)z^2-\frac{2iJ}{E}z-1=0,
\end{eqnarray}
can be easily solved numerically.

Our findings for the random-field models (\ref{2.01}), (\ref{5.01}) with $\Gamma=0.1$
are shown by thick lines in the two lower panels of Figs.~\ref{fig2}, \ref{fig3}, \ref{fig6}, \ref{fig7}
and in Figs.~\ref{fig4}, \ref{fig8}.
From the analysis of nonrandom models in Secs.~\ref{sec3} and \ref{sec4}
we know that essential enhancement of MCE occurs around QCPs and QTPs.
From the reported results for the random-field chains 
(compare, e.g., thin and thick lines in Figs.~\ref{fig4} and \ref{fig8})
we conclude 
that just around these special points the MCE is extremely sensitive to randomness.
As can be seen in Figs.~\ref{fig4} and \ref{fig8},
even small randomness leads to a rounding of sparks in the dependence $\Gamma_h(h)$
and noticeable diminishing of maximal values of $\Gamma_{h}$.

\section{Conclusions}
\label{sec6}

In this work
we have studied the MCE for the spin-1/2 $XX$ chain
with the three-spin interactions of the $XZX+YZY$ and $XZY-YZX$ types.
The considered models have more parameters
(in addition to the magnetic field we can also vary the strength of three-spin interactions),
they contain several lines of QCPs and QTPs, 
and manipulation with MCE becomes possible.
We have found that the quantum phase transition lines clearly manifest themselves
for the MCE in the low-$s$ or low-$T$ regimes.
The ground-state phase diagrams can be perfectly reproduced by measuring of $\Gamma_h$ in the limit $T\to 0$.
The vicinity of QCPs or QTPs is very effective for cooling
since low temperatures are achieved by only small decrease of the field.
Particularly strong variation of temperature (at $s={\rm{const}}$) or of entropy (at $T={\rm{const}}$)
with varying the magnetic field occurs in the vicinity of a QTP.

We have discussed the MCE in a random quantum spin chain.
We have found that even small randomness can noticeably diminish 
an enhanced MCE in proximity to a QCP/QTP. 

The considered models, thanks to their simplicity, have enabled the rigorous analysis 
of the thermodynamic quantities of interest.
Although we do not know any particular compound which can be described by the studied models,
our results might have a general merit 
being useful for understanding the effects of proximity to QTP and randomness on the MCE.
On the other hand, with further progress in material sciences and synthesis of new magnetic chain compounds,
the lack of experimental data and comparison between theory and experiment may be resolved in future.

\section*{Acknowledgments}

The authors thank 
A.~Honecker,
A.~Kl\"{u}mper 
J.~Richter,
J.~Sirker,
T.~Vekua,
and 
M.~E.~Zhitomirsky
for useful comments.
O. D. and V. O. acknowledge financial support 
of the organizers 
of the SFB602 workshop on Localized excitations in flat-band models 
(G\"{o}ttingen, April 12-15, 2012)
and
of the 504. WE-Heraeus-Seminar on ``Quantum Magnetism in Low Spatial Dimensions''
(Bad Honnef, 16-18 April 2012)
where the paper was finalized.
V. O. expresses his gratitude 
to the Department of Theoretical Physics of the Georg-August University (G\"{o}ttingen) 
and ICTP (Trieste) 
for warm hospitality during the work on this paper. 
He also acknowledges financial support 
from DFG (Grant No. HO 2325/8-1), 
ANSEF (Grant No. 2497-PS), 
Volks\-wagen Foundation (Grant No.\ I/84 496),
SCS-BFBR 11RB-001 grant,
and joint grant of CRDF-NFSAT and the State Committee of Science of Republic of Armenia 
(Grant No. ECSP-09-94-SASP).


\begin{thebibliography}{99}
\bibitem{footnote}
We may also estimate $\Gamma_h(h)$ in the limit $T\to 0$ within a spin-liquid phase.
For this we have to take into account 
that the equation $\varepsilon_k=0$ has two ($K<1/2$), three ($K=1/2$), or four ($K>1/2$) solutions $\{k_i\}$
and $\varepsilon_k=\varepsilon_i^{(1)}(k-k_i)+\varepsilon_i^{(2)}(k-k_i)^2/2!+\ldots$ around $k_i$.
Further we have to approximate $n_k(1-n_k)$ in the spirit of Eq.~(\ref{3.09})
and after extending the limits of integration to evaluate relevant integrals in Eq.~(\ref{2.07}).
The final result reads:
\begin{eqnarray}
\Gamma_h
\to
-\frac{\sum_i\varepsilon_i^{(2)}/\left\vert\varepsilon_i^{(1)}\right\vert^3}{\sum_i 1/\left\vert\varepsilon_i^{(1)}\right\vert}.
\nonumber
\end{eqnarray}
Alternatively, we may begin with $\Gamma_h$ given in Eq.~(\ref{2.09}).
As was mentioned in Sec.~\ref{sec2}, $1/{\rm{ch}}^2[\omega/(2T)]$ tends to $4T\delta(\omega)$ in the limit $T\to 0$
and hence only a small region around $\omega=0$ is relevant.
Expanding $\rho(\omega)$ around $\omega=0$,
$\rho(\omega)=\rho(0)+\rho^{(1)}(0)\omega +\ldots$,
we arrive at
\begin{eqnarray}
\Gamma_h
\to
- \frac{d\rho(\omega)/d\omega\vert_{\omega=0}}{\rho(\omega)\vert_{\omega=0}}.
\nonumber
\end{eqnarray}
These two results for $\Gamma_h$ are equivalent.
\end{thebibliography}
\end{document}